\documentclass[preprint]{aastex}
\newcommand{\abs}[1]{\left|#1\right|} 

\begin{document}

\title{Studying the Galactic Bulge Through Spectroscopy of Microlensed
Sources: I. Theoretical Considerations}

\author{Stephen R. Kane\altaffilmark{1} and Kailash C. Sahu}
\affil{Space Telescope Science Institute, 3700 San Martin Drive,
Baltimore, MD 21218, U.S.A.}
\email{skane@aastro.ufl.edu, ksahu@stsci.edu}

\altaffiltext{1}{Currently at the Department of Astronomy, University
of Florida, 211 Bryant Space Science Center, Gainesville, FL 32611-2055,
U.S.A.}

\begin{abstract}

The observed spectra of the microlensed sources towards the Galactic
bulge may be used as a tool for studying the kinematics and extinction
effects in the Galactic bulge. In this paper, we first investigate the
expected distribution of the microlensed sources as a function of depth
within the Galactic bulge. Our analysis takes a magnitude limited
microlensing survey into account, and includes the effects of
extinction. We show that, in the current magnitude limited surveys, the
probability that the source lies at the far side of the bulge is larger
than the probability that the source lies at the near side. We then
investigate the effects of extinction on the observed spectra of
microlensed sources. Kurucz model spectra and the observed extinctions
towards the Galactic bulge have been used to demonstrate that the
microlensed sources should clearly show the effects of extinction which,
in turn, can be used as a statistical measure of the contribution of
the disk lenses and bulge lenses at different depths. The spectra of the
microlensed sources provide a unique probe to derive the radial
velocities of a sample which lies preferentially at the far side of the
Galactic bulge. The radial velocities, coupled with the microlensing
time scales, can thus be useful in studying the 3-dimensional
kinematics of the Galactic bulge.

\end{abstract}

\keywords{Galaxy: structure --- Galaxy: stellar content ---
stars: kinematics}

\section{INTRODUCTION}

The topic of gravitational microlensing has experienced a dramatic
increase in scientific interest over recent years. This has been largely
due to the realization of its wide-ranging applications, such as the
detection of planets and the study of Galactic structure. More than 400
microlensing events have now been detected toward the Galactic bulge and
Magellanic Clouds by the microlensing survey teams EROS, MACHO, OGLE, DUO,
and MOA. The contribution of the bulge stars to the observed microlensing
optical depth towards the Galactic bulge was estimated to be approximately
60\%, the other 40\% being due to the stars in the Galactic disk
\citep{kir94}. The optical depth to gravitational microlensing, as
estimated from the observed microlensing events by OGLE \citep{uda94},
MACHO \citep{alc97}, EROS \citet{afo03}, and MOA \citet{sum03}, is
significantly larger than the values predicted by theoretical models
\citep{pac91}. It has been argued \citep{pac94} that the observed optical
depth can be best explained if the effect of the Galactic bar and its
inclination are correctly taken into account. A consequence of the fact
that a large fraction of the events are due to bulge-bulge lensing is, as
we will show in more detail later, that the lensed stars will
preferentially be located on the far side of the bulge in order that there
be sufficient stars along the line of sight to cause microlensing. It was
suggested by \citet{sta95} that this would mean that there should be a
systematic offset in the apparent magnitude between observed stars and
lensed stars. We present a similar discussion by examining the model
spectra of different spectral classes and investigating the effects of
extinction on stars located on the far side of the bulge.

We first calculate the contributions of the different layers in the
bulge to the microlensing optical depth, as has previously been
calculated by \citet{kir94} and \citet{zha95}. We have, however,
considered the additional effect of extinction in some detail. We then
proceed with a discussion on how the spectra of the microlensed sources
can be used as a measure of the extinction which, in turn, can be used
as a statistical measure of the contribution of the disk lenses and
bulge lenses at different depths. We show that the spectra of the
microlensed sources can be a useful probe for the 3-dimensional
structure of the Galactic bulge. We have also undertaken such a
spectroscopic study of the microlensed sources, the details of which
are the subject of another paper \citep{kan03}.

\section{CALCULATING THE PROBABILITY OF MICROLENSING}

The aim here is to calculate the microlensing probability for sources at
various depths within the Galactic bulge and to see how the microlensing
optical depth varies for sources at various depths taking the effect of
extinction into account. We first develop the theory for a general
distribution of microlensing sources and will later restrict the source
distribution to those located in the Galactic bulge. For these
calculations, the formalism used by Sahu \citep{sah94a,sah94b} shall be
adopted.

Firstly, the number of sources observed at various depths shall be
estimated. As we will discuss in more detail later, the intrinsic
distribution of the Galactic bulge sources can be expressed by an
exponential distribution, the density being maximum at the central part
and falling on either side. The extinction and the distance modulus
increase as one goes deeper into the bulge, and both these effects make
the sources fainter. So, in a magnitude limited survey (such as the
current ones), the combined effect of the extinction and the distance
is a monotonous decrease in the observed number of stars as we go from
the nearest to the farthest region along a given line of sight, which
needs to be folded in with the intrinsic distribution of the sources in
order to obtain their observed distribution.

To express the above effects, let us assume that the ratio of the
stellar number density at the central part of the bulge to the stellar
density in the near-side of the bulge is $a_1$ and the ratio of the
stellar number density at the central part of the bulge to the stellar
density in the far-side of the bulge is $a_2$. Thus, $a_1$ and $a_2$
include the effects of localized extinctions and the intrinsic
distributions.

Let $N_{tot}$ be the number of stars being monitored in a region.
Assuming the extinction to be uniform in depth within the bulge, the
number of observed stars at any layer $dx$, at a depth of $x$ (as
measured from the central plane of the Galactic bulge, see Figure 1),
may be expressed as
\begin{equation}
  N_{obs}(x) = \left\{
  \begin{array}{ll} N_0
    a_1^{-x/d} & \mbox{if $D_s < R_0$}\\
    N_0 a_2^{-x/d} & \mbox{if $D_s > R_0$}
  \end{array}
  \right.
\end{equation}
where $N_0$ is a constant and represents the observed number
of stars per unit depth in the central region of the Galactic bulge and
$d$ is the depth to which the bulge extends on either side from the
center. $R_0$ is the distance to the central region of the bulge and
$D_s$ is the distance to the source (i.e., the distance from the
observer to the plane corresponding to $x$ so that $x = {\abs{D_s-R_0}}$).

Integrating this expression along the line of sight, we can write
\begin{eqnarray}
  N_{tot} & = & \int_{0}^{d} N_0 a_1^{-x/d} \, dx +
  \int_{0}^{d} N_0 a_2^{-x/d} \, dx \nonumber\\
  & = & N_0 \left[ \frac{d}{\ln a_1} \left( 1 - \frac{1}{a_1} \right) +
    \frac{d}{\ln a_2} \left( 1 - \frac{1}{a_2} \right) \right]
\end{eqnarray}

The fraction of area covered by the Einstein rings of all the individual
stars lying in front of a source at distance $D_s$ may be expressed as
\begin{equation}
  A_f(D_s) = \int_0^{D_s} \pi R_E^2(D_d) n(D_d) \, dD_d
\end{equation}
where $D_d$ is the distance to the lens and $n(D_d)$ is the stellar
number density at distance $D_d$. If we define $\rho(D_d)$ as the
average mass density at depth $D_d$ then we see that $n(D_d) = \rho(D_d)
/ M$. Since the Einstein ring radius for a lens of mass $M$ in this
case is given by
\begin{equation}
  R_E = \sqrt{\frac{4 G M}{c^2} \frac{D_d (D_s-D_d)}{D_s}}
\end{equation}
Equation (3) can be rewritten as
\begin{equation}
  A_f(D_s) = \frac{4 \pi G}{c^2} \int_0^{D_s} \rho(D_d)
\frac{D_d (D_s-D_d)}{D_s} \, dD_d
\end{equation}

The instantaneous probability that an observed microlensed star is at
a given distance $D_s$,
can be written as
\begin{equation}
  p = \frac{A_f(D_s) N_{obs}(D_s)}{N_{tot}}
\end{equation}
Note that, in our desire to treat both the source and lens distributions
in a similar manner, we have chosen our expression for the number of
observed stars at $D_s$ to be $N_{obs}(D_s) \, dD_s$, and not $N_{obs}(D_s)
\, dD_s \, d\Omega$. This would suggest that $N_{obs}(D_s)$ may increase
with $D_s$ since the solid angle $d\Omega \propto D_s^2$. On the other
hand, in a magnitude-limited sample, the $N_{obs}(D_s)$ would decrease with
$D_s$ since the observed flux decreases as $D_s^{-2}$. The above two effects
tend to cancel each other, but not exactly, particularly when there is
extinction. In our general case of the exponential distribution for the
bulge (section 2.3), we have chosen $a_1$ and $a_2$ to be different, which
should absorb this extra effect.

The total microlensing optical depth $\tau$ can now be calculated. If
$N_{tot}$ stars within the bulge are being monitored, then the
instantaneous probability of observing a lensing event with an
amplification of $A > 1.34$ is given by
\begin{equation}
  P = \int_{b_n}^{b_f} N_{obs}(D_s) A_f(D_s) \, dD_s
\end{equation}
where we assume all the monitored sources to be within the Galactic
bulge and, for convenience, we have defined $b_n = R_0 - d$ and $b_f =
R_0 + d$ (i.e., the distances from the observer to the near side and far
side of the bulge respectively).

The optical depth for microlensing, which is the instantaneous
probability that a given star is microlensed, can be written as
\begin{equation}
  \tau = P / N_{tot}
\end{equation}

Assuming a uniform extinction with depth is a simple model with some
inherent uncertainties and is unlikely to be representative of the real
extinction in the line of sight to microlensed sources. However, this
does not affect the conclusions of this analysis. A full modeling of the
extinction has its own uncertainties and is beyond the scope of this
work.

It should be noted that in our calculations we are not comparing
extinctions at different parts in the bulge but rather we are comparing
the differential extinctions between microlensed and non-microlensed
sources in the same part of the sky. It is know that the dust layer
when looking towards the bulge is patchy and predominantly in the disk
\citep{arp65}. However, this will not significantly affect our
calculations as the angular distance between the microlensed source and
the corresponding lens will be small enough such that the foreground
extinction may be considered to be the same for both objects \citep{zha00}.

We will carry out these calculations using 3 different approximations:
(i) constant density of stars between the observer and the source,
(ii) constant (but different) densities for the disk and the bulge, and
(iii) an exponential density model for the bulge. The last approximation
is likely to be the best model, and hence the results of this model will
be used in our forthcoming spectral analysis \citep{kan03}. For
completeness however, we present the calculations in the other two
approximations, since these calculations may be useful for other lines
of sight, including the lines of sight far from the Galactic Center, and
possibly towards the LMC and M31.

\subsection{Constant Density Between Observer and Source}

First, let us assume a constant density of matter for the lenses between
the observer and the source and make the substitution $y = D_d / D_s$,
then equation (5) becomes
\begin{eqnarray}
  A_f(D_s) & = & \frac{4 \pi G D_s^2}{c^2} \int_0^1 \rho(D_d) y (1-y)
  \, dy \nonumber\\
  & = & \frac{2 \pi G \rho D_s^2}{3 c^2}
\end{eqnarray}

Since the matter density in the line of sight is constant, extinction is
the only effect which changes the observed number density of sources in
the line of sight. Let us define the intrinsic stellar density $N_n = N_0
a$ where $N_0$ is the observed stellar density at the mid-plane along the
line of sight. Let $d_c = 2 d$ be the total depth, and $a_c = a^2$ which
corresponds to the ratio of the stellar density from the near to the far
side. In this case, equation (1) takes the form
\begin{equation}
  N_{obs}(D_s) = N_n a_c^{-D_s/d_c}
\end{equation}
Integrating this expression yields
\begin{equation}
  N_{tot} = \int_0^{d_c} N_n a_c^{-D_s/d_c} \, dD_s = N_n d_c \left(
  \frac{1 - 1/a_c}{\ln a_c} \right)
\end{equation}

Substituting equations (9-11) into equation (6) yields
\begin{equation}
  p = \frac{2 \pi G \rho D_s^2}{3 c^2 d_c} \left( \frac{\ln a_c}{1 - 1/a_c}
  \right) a_c^{-D_s/d_c}
\end{equation}
This probability $p$ is plotted in Figure 2 as a function of
$D_s$ assuming a density of $\rho = 0.20 \, M_\odot
\mathrm{pc}^{-3}$.

As seen in Figure 2, for $a_c = 1$ (which corresponds to no extinction), the
probability that a given observed microlensed source is at a distance
$D_s$ increases monotonically with $D_s$. It is clear that $A_f(D_s)$,
the fractional area covered by the lenses for a source at distance $D_s$,
increases monotonically with distance. However, for $a_c > 1$, the number
of observed stars decreases with distance because of extinction. As seen
in equation (6), $p$ is a multiplication of these two quantities which, as
demonstrated in Figure 2, increases steeply with distance for $a_c =1$,
and increases less steeply with distance for increasing values of $a_c$.
This is also demonstrated in Table 1 which gives the probabilities that a
given microlensed star is at distances $D_s = 7.0 \, \mathrm{kpc}$,
$D_s = 8.0 \, \mathrm{kpc}$, and $D_s = 9.0 \, \mathrm{kpc}$ for various
values of $a_c$. It is worth reiterating
here that $p = [A_f(D_s) N_{0bs}(D_s) / N_{tot}]$ is the probability that
a given observed microlensed source is at a distance $D_s$, which is
distinct from the probability that a given observed star at the distance
$D_s$ is microlensed; the latter simply scales as the microlensing optical
depth $\tau$ which monotonically increases with the distance.

Substituting equation (9) and equation (11) into equation (7) gives
\begin{equation}
  P = \frac{2 \pi G \rho}{3 c^2} \frac{N_{tot}}{d_c} \left(
  \frac{\ln a_c}{1 - 1/a_c} \right) \int_0^{d_c} a_c^{-D_s/d_c} D_s^2 \,
  dD_s
\end{equation}
where the integrand can be solved analytically via repeated
integration by parts. Dividing the result by $N_{tot}$ yields an
expression for the optical depth for gravitational microlensing
\begin{equation}
  \tau = \frac{2 \pi G \rho d_c^2}{3 c^2} \frac{a_c}{a_c-1} \left(
  \frac{2}{(\ln a_c)^2} - \frac{1}{a_c} - \frac{2}{a_c \ln a_c}
  - \frac{2}{a_c (\ln a_c)^2} \right)
\end{equation}

The value of $\tau$ as given in equation (14) is plotted in Figure 3 as a
function of $a_c$, which shows the variation of optical depth with
increasing extinction. As seen from this figure, the optical depth in this
case is close to $5 \times 10^{-7}$ for a large range of $a_c$ since the
intrinsic matter distribution, which plays the major role in the
determination of optical depth, is constant.

However, the assumption of constant density along the line of sight
may be valid for microlensed sources in the Galactic disk (e.g.,
OGLE-1999-CAR-01 \citep{uda00}), but can not be applied to sources
in the Galactic bulge.

\subsection{Constant Density for Disk and Bulge}

If we assume a constant density of matter for the lenses in the disk
and the bulge, given by $\rho_d$ and $\rho_b$, respectively, equation
(5) becomes
\begin{equation}
  A_f(D_s) = \frac{4 \pi G}{c^2} \left( \int_0^{b_n} \rho_d \frac{D_d
    (D_s-D_d)}{D_s} \, dD_d + \int_{b_n}^{D_s} \rho_b \frac{D_d
    (D_s-D_d)}{D_s} \, dD_d \right)
\end{equation}

Since the optical depth of microlensing of disk stars is relatively
small \citep{kir94}, we have assumed that all the sources are within
the Galactic bulge, i.e., $b_n < D_s < b_f$.

Making the substitution $y = D_d / D_s$, equation (15) simplifies to
\begin{equation}
  A_f(D_s) = \frac{4 \pi G D_s^2}{c^2} \left( \rho_d \int_0^{b_n/D_s}
  y (1-y) \, dy + \rho_b \int_{b_n/D_s}^1 y (1-y) \, dy \right)
\end{equation}
Solving this yields the following expression for the fractional area
covered by the Einstein rings, which is the same as the probability
that a given star at a distance $D_s$ is microlensed at any given
time:
\begin{equation}
  A_f(D_s) = \frac{4 \pi G}{c^2} \left( \rho_d \left( \frac{b_n^2}{2}
  - \frac{b_n^3}{3 D_s} \right) + \rho_b \left( \frac{D_s^2}{6} -
  \frac{b_n^2}{2} + \frac{b_n^3}{3 D_s} \right) \right)
\end{equation}
The first term in equation (17) is the contribution of the disk lenses
and the second term is that of the bulge lenses.

Since the intrinsic number density of sources within the bulge is
constant in this case, extinction is the only effect that changes the
observed number density of sources in the line of sight. As in the
previous case, let us define the observed stellar density at the near
side $N_n = N_0 a$ where $N_0$ is the observed stellar density at the
mid-plane of the bulge. Let $d_c = 2 d$ be the total depth of the bulge,
and $a_c = a^2$ which corresponds to the ratio of the stellar density
from the near to the far side of the bulge. In this case, equation (1)
takes the form

\begin{equation}
  N_{obs}(x_c) = N_n a_c^{-x_c/d_c}
\end{equation}
where $x_c$ ($= D_s - b_n$) is the depth as measured from the near side
from the bulge. Integrating this expression yields
\begin{equation}
  N_{tot} = \int_0^{d_c} N_n a_c^{-x_c/d_c} \, dx_c = N_n d_c \left(
  \frac{1 - 1/a_c}{\ln a_c} \right)
\end{equation}

Let $R_0 = 8.0 \, \mathrm{kpc}$, $b_n = 7.0\, \mathrm{kpc}$, $d_c = 2.0 \,
\mathrm{kpc}$, $\rho_d = 0.07 \, M_\odot \mathrm{pc}^{-3}$, and $\rho_b = 0.50
\, M_\odot \mathrm{pc}^{-3}$. Substituting equation (17-19) into
equation (6) yields the instantaneous probability that an observed star at a
given distance $D_s$ is microlensed

\begin{equation}
  p = \frac{4 \pi G}{c^2 d_c} \left( \frac{\ln a_c}{1-1/a_c} \right)
  a_c^{-x_c/d_c}  \left( \rho_d \left( \frac{b_n^2}{2} -
  \frac{b_n^3}{3 D_s} \right) + \rho_b \left( \frac{D_s^2}{6} -
  \frac{b_n^2}{2} + \frac{b_n^3}{3 D_s} \right) \right)
\end{equation}
This probability $p$ is plotted in Figure 4 as a function of $D_s$
using the above parameters for the bulge. 

Figure 4 shows that, when a constant density is assumed for the bulge, the
probability that a given microlensed source (in an observed sample of stars)
is on the far side of the bulge quickly becomes dominated by the level of
extinction. Indeed, as shown in Table 2, the probability that a given
microlensed star is located at the near side of the bulge ($D_s = 7.0 \,
\mathrm{kpc}$) is larger than the probability that that a microlensed source
is located at the far side of the bulge ($D_s = 9.0 \, \mathrm{kpc}$) for
$a_c \gtrsim 2$, simply because too few of the sample of obseerved stars are
at 9 kpc. This is analogous to the previous case, where the reasons have been
explained in detail.

For completeness, the total microlensing optical depth will now be calculated.
Substituting equations (17-19) into equation (7) yields an expression for the
optical depth for gravitational microlensing
\begin{equation}
  \tau = \frac{4 \pi G}{c^2 d_c} \left( \frac{\ln a_c}{1 - 1/a_c} \right)
  \int_{b_n}^{b_f} \left( (\rho_b - \rho_d) a_c^{-x_c/d_c}
  \left( \frac{b_n^3}{3 D_s} - \frac{b_n^2}{2} \right) +
  \frac{\rho_b}{6} a_c^{-x_c/d_c} D_s^2 \right) \, dD_s
\end{equation}

The value of $\tau$ as given in equation (21) is plotted in Figure 5 as a
function of $a_c$, which shows the variation of optical depth with increasing
extinction.

For $a_c = 3.0$ the optical depth is $\approx 6.76 \times 10^{-7}$ and for
$a_c = 7.0$ the optical depth is $\approx 5.16 \times 10^{-7}$. These values
are slightly lower than an optical depth of $\approx 8.5 \times 10^{-7}$
estimated by \citet{kir94} using a similar Galactic model but without taking
extinction into account. This is the result one would expect when extinction
is added to the optical depth calculations. However, our calculation is
dependent upon the choice of $\rho_d$ and $\rho_b$ which, given the nature
of the model, will be crude estimates at best.

\subsection{Exponential Density Model for Bulge}

A more realistic approximation for the density distribution of lenses
between the observer and the source is to use an exponential-type
function to represent the density profile over the Galactic bulge which
includes the disk component. The model used to do this is a simplified
version of the E2 triaxial Galactic bulge model fitted by \citet{dwe95}
with an exponential cutoff
at 2.4 kpc. This model is defined as
\begin{equation}
  \rho_{E2}(i,j,k) = \rho_0 \exp (-r)
\end{equation}
where
\begin{equation}
  r = \left[ \left( \frac{i}{i_0} \right)^2 + \left( \frac{j}{j_0}
    \right)^2 + \left( \frac{k}{k_0} \right)^2 \right]^{1/2}
\end{equation}
is the distance from the Galactic centre. Treating this from a purely
radial perspective and centering the coordinate system on the observer
results in the density profile given by
\begin{equation}
  \rho(D_d) = \rho_0 \exp \left[ - \frac{|R_0 - D_d|}{i_0} \right]
\end{equation}
where the normalisation constant is given by
\begin{equation}
  \rho_0 = \frac{M}{8 \pi i_0 j_0 k_0}
\end{equation}
The value of $\rho_0$ shall be calculated by adopting the total bulge
mass estimated by \citet{dwe95} of $M = 1.3 \times 10^{10} \, M_\odot$
and using the scale lengths of $i_0 = 0.71$, $j_0 = 0.18$, and $k_0 =
0.25 \, \mathrm{kpc}$ for a Galactocentric distance of $R_0 = 8.0 \,
\mathrm{kpc}$. This model is valid for a line of sight that passes
approximately through the Galactic center $(l,b) = (0,0)$. Although it
would be interesting to conduct a thorough analysis of a line of sight
which passes through Baade's window, $(l,b) = (1,-4)$, changing the
direction slightly will alter the stellar density profile but the
extinction model will remain largely the same and hence so will the
results.

Using this model, equation (5) becomes
\begin{equation}
  A_f(D_s) = \frac{4 \pi G \rho_0}{c^2} \int_0^{D_s} \exp \left[ -
    \frac{|R_0 - D_d|}{i_0} \right] \frac{D_d (D_s-D_d)}{D_s} \, dD_d
\end{equation}
Solving this equation (see Appendix) yields the following expression for
the fractional area covered by the Einstein rings:
\begin{equation}
  A_f(D_s) = \left\{
  \begin{array}{ll}
    \frac{4 \pi G \rho_0}{c^2} \frac{i_0^2 \alpha}{D_s}
    & \mbox{for $D_s < R_0$}\\
    \frac{4 \pi G \rho_0}{c^2} \frac{i_0 (2 i_0^2 \beta - 2 R_0^2 + D_s
      \gamma)}{D_s}
    & \mbox{for $D_s > R_0$}
  \end{array} \right.
\end{equation}
where
\begin{eqnarray}
  \alpha & = & e^{-R_0/i_0} (D_s (1 + e^{D_s/i_0}) + 2 i_0 (1 -
  e^{D_s/i_0})) \nonumber\\
  \beta & = & e^{-(D_s-R_0)/i_0} + e^{-R_0/i_0} -2 \nonumber\\
  \gamma & = & i_0 e^{-(D_s-R_0)/i_0} + i_0 e^{-R_0/i_0} + 2 R_0 \nonumber
\end{eqnarray}

Following equations (1) and (2), we can express the observed density of
sources as
\begin{equation}
  N_{obs}(D_s) = \left\{
  \begin{array}{ll}
    N_0 a_1^{-(R_0-D_s)/d} & \mbox{if $D_s < R_0$}\\
    N_0 a_2^{-(D_s-R_0/d} & \mbox{if $D_s > R_0$}
  \end{array}
  \right.
\end{equation}
where $N_0$ is the the observed number of stars per unit depth at the
mid-plane of the bulge. We have assumed the lens distribution to be
distinct from the source distribution within the bulge since the source
density distribution has an extra dependence on the sensitivity of
observations and the extinction. Note that when $a_1$ and $a_2$ are
greater than unity, the observed source density peaks at the mid-plane,
which is a realistic case for the Galactic bulge. Also, the definitions
of $a_1$ and $a_2$ require that $a_1 < a_2$ since the extinction reduces
the intrinsic density contrast between the center and the near side of
the bulge, whereas it increases the density contrast between the center
and the far side of the bulge. As before, $d$ is the depth to which the
bulge extends on either side from the center. 

Integrating this expression along the line of sight as in equation
(2), we can write the total number of observed sources
\begin{eqnarray}
  N_{tot} & = & 
  N_0 \left[ \frac{d}{\ln a_1} \left( 1 - \frac{1}{a_1} \right) +
    \frac{d}{\ln a_2} \left( 1 - \frac{1}{a_2} \right) \right] \nonumber
\end{eqnarray} 
or alternatively
\begin{eqnarray}
  N_{tot} & = & 
  N_0 d \left[ \frac{(a_1-1) a_2 \ln a_2 + (a_2-1) a_1 \ln a_1}{a_1
      a_2 \ln a_1 \ln a_2}  \right]
\end{eqnarray}

Substituting equations (27-29) into equation (6), we can write the
instantaneous probability that an observed microlensed star is at a
given distance $D_s$ 
\begin{equation}
  p = \left\{
  \begin{array}{ll}
    \frac{4 \pi G \rho_0}{c^2 d} 
    \left[ \frac{a_1 a_2 \ln a_1 \ln a_2} {(a_1-1) a_2 \ln a_2 + (a_2-1)
	a_1 \ln a_1} \right] a_1^{-(R_0-D_s)/d} \left( \frac{i_0^2
      \alpha}{D_s} \right)
    & \mbox{for $D_s < R_0$}\\
    \frac{4 \pi G \rho_0}{c^2 d} 
    \left[ \frac{a_1 a_2 \ln a_1 \ln a_2} {(a_1-1) a_2 \ln a_2 + (a_2-1)
	a_1 \ln a_1} \right] a_2^{-(D_s-R_0)/d} \left( \frac{i_0 (2 i_0^2
      \beta - 2 R_0^2 + D_s \gamma)}{D_s} \right)
    & \mbox{for $D_s > R_0$}
  \end{array} \right.
\end{equation} 
where we have assumed that all the sources are within the Galactic bulge,
i.e., $b_n < D_s < b_f$.
The probability $p$ is plotted in Figure 6 as a function of $D_s$ using
the above model.

Figure 6 shows that an exponential density model dramatically affects the
probability that a microlensed source is at a distance $D_s$. For likely
values of $a_1 \lesssim 2$ and $a_2 \lesssim 2$, the probability of a
microlensed source being in the far-side of the bulge is higher. For higher
values of $a_1$ and $a_2$, the probability peaks at the central plane. This is
also seen from Table 3 which gives the probabilities that a given microlensed
star is at distances $D_s = 7.0 \, \mathrm{kpc}$, $D_s = 8.0 \, \mathrm{kpc}$,
and $D_s = 9.0 \, \mathrm{kpc}$ for $a_1 = 1$. The table shows that for this
particular case of $a_1 = 1$, the probability that a microlensed star is on
the far side of the bulge is larger than the probability that it is on the
near side of the bulge if $a_2 \lesssim 3$.

The total microlensing optical depth can now be calculated. Substituting
equation (7) and equations (28-30) into equation (8) yields an expression
for the optical depth for gravitational microlensing
\begin{eqnarray}
  \tau & = & \frac{4 \pi G \rho_0}{c^2 d} 
  \left[ \frac{a_1 a_2 \ln a_1 \ln a_2} {(a_1-1) a_2 \ln a_2 + (a_2-1) a_1
      \ln a_1} \right]  \nonumber\\
  & & \left[ i_0^2 \int_{b_n}^{R_0} a_1^{-(R_0-D_s)/d} \frac{\alpha}{D_s} \,
    dD_s + i_0 \int_{R_0}^{b_f} a_2^{-(D_s-R_0)/d} \frac{2 i_0^2 \beta - 2
      R_0^2 + D_s \gamma}{D_s} \, dD_s \right]
\end{eqnarray}

The value of the microlensing optical depth $\tau$ as given by equation (31)
is plotted in Figure 7 as a function of $a_2$, for various values of $a_1$
ranging from 1 to 5. As seen from the figure, the microlensing optical
depth increases with $a_1$ and decreases with $a_2$. For realistic values of
$a_1 = 2.0$ and $a_2 = 3.0$, the optical depth is  $4.50 \times 10^{-6}$, and
for $a_1 = 2.0$ and $a_2 = 2.0$, the optical depth is $4.84 \times 10^{-6}$.
These estimates for the microlensing optical depth towards the Galactic
bulge are comparable to the values of $(3.3 \pm 1.2) \times 10^{-6}$ and
$(3.9_{-1.2}^{+1.8}) \times 10^{-6}$ measured by \citet{uda94} and
\citet{alc97} respectively. \citet{alc00a} derive a value of $\tau_{total} =
(2.43_{-0.38}^{+0.39}) \times 10^{-6}$ from difference image analysis, and
\citet{pop00} derive a value of $\tau = (2.43\pm{0.4}) \times 10^{-6}$,
which are also consistent with our estimated value. \citet{zha95} and
\citet{pac94} have estimated a similar value of $(2.2 \pm 0.45) \times
10^{-6}$ and $(3.3 \pm 1.2) \times 10^{-6}$, respectively, by taking the
inclination of the bar into account.

The limiting apparent magnitude of the current surveys is about $m_V = 20$
which at the near side of the bulge would correspond to an absolute magnitude
of $M_V = 5.8$ and an absolute magnitude of $M_V = 5.3$ at the far side of the
bulge were there no extinction. However, the estimated foreground extinction
towards Baade's window has been estimated by \citet{sta96} to lie in the range
of 1.26 to 2.79 magnitudes, depending on the line of sight. Hence the limiting
absolute magnitudes of the stars in this direction lies is in the range $M_V$ =
4.5 to 2.5 (plus an additional 0.5 magnitudes because of the larger  distance).
If the distribution of stars amongst different spectral types is assumed to be
similar to what is observed in the solar neighbourhood \citep{wie83,kro93} then
the stellar number density is relatively flat in this region. This distribution
shows that the ratio in the observed stellar number density from the center to
the near or far side of the bulge is between $\sim 1.3$ to 2.8, depending on
the internal extinction.

For $a_1 = 2.0$ and $a_2 = 3.0$, the probability that an observed microlensed
source is at a distance $D_s = 9$ kpc is $\approx 8$ times larger than the
probability that it is at a distance $D_s = 7$ kpc. For $a_1 = 2.0$ and $a_2
= 2.0$, the probability that an observed microlensed source is at a distance
$D_s = 9$ kpc is also $\approx 13$ times larger than the probability that it
is at a distance $D_s = 7$ kpc, and the full probability distribution as a
function of $D_s$ is shown in Figure 6. This shows that, on average, stars
that are lensed in Baade's window are more likely to be at the far side of
the bulge than the near side for reasonable values of $a_1$ and $a_2$.

If lensed stars are indeed predominantly on the far side of the bulge
then this provides us with a useful tool for studying the Galactic
structure at the far side of the bulge. Radial velocity measurements
from spectra of microlensed sources combined with the measured time
scale of the events may be used as a unique probe into the
3-dimensional kinematics of the far side of the bulge.

\subsection{Effect of the Galactic Bar}

We have so far neglected the effect of the Galactic bar. The Galactic
bar, however, is known to exist both from infrared and microlensing
observations \citep{pac94,ger00,ham00}. A number of photometric and
dynamic indications also point out the presence of the Galactic bar
\citep{bin91,whi91,wei92}. The effect of the bar is not only obvious
from the microlensing observations, but its inclination might be
essential to account for the optical depth observed in at least some
of the lines of sight (also see \citet{bin00}). Furthermore, the
inclination of the major axis of the bar with respect to the line of
sight, as independently determined from the microlensing observations,
is consistent with the inclination of $\sim 15$ degrees determined
from the earlier infrared observations. So a natural question to ask
is, how does the presence of the Galactic bar effect the analysis
presented here? In particular, what is the effect of the Galactic bar
on the spectroscopic observations, both in terms of extinction and
kinematics?

The effect of the Galactic bar is expected to be seen even more clearly
in the spectroscopic observations. The reason is two-fold. The first
part has to do with the extinction. A good part of the Baade's window
is thought to have little extinction. The Galactic bar on the other
hand, which clearly occupies only a part of the region surveyed for
microlensing (cf. \citet{bli91,uda94,alc97}), is bright in the
far-infrared wavelengths as seen in the IRAS SKYFLUX maps \citep{bei85},
clearly indicating the presence of cool dust. In such a case, the
microlensed sources in the region of the Galactic bar should certainly
show larger extinction compared to the unlensed sources. The second part
has to do with the expected kinematics. Although there is some
uncertainty in the models, the objects in the Galactic bar, in general,
are expected to be kinematically distinct from those in the outer parts
of the bulge (see, for example, \citet{ken87,bli91}). Thus the effect of
the bar may be more prominently seen in the kinematic distribution.
Indeed, the kinematics of the microlensed sources may provide very
meaningful constraints on the structure of the bar. In order to see this
extra effect of the Galactic bar, however, the observations must be done
for the sources in a restricted region covering the Galactic bar. 

\subsection{Blending Effects}

The Galactic bulge fields are crowded in general. As a result, many of
the microlensed sources may be blended with other stars, and this effect
must be taken into account in estimating the extinctions and the radial
velocities of the microlensed sources from the observed spectra. 
(Note that the blending has no effect on the theoretical estimate of
the optical depth described above. The discussion here refers only
to the spectroscopic observations and role of blending in such
observations). As the recent HST images towards the LMC and the Galactic
bulge show, blending can be an important effect. In the case of the
microlensing events towards the LMC, each `microlensed star' typically
splits into 2 or more sources in the high spatial-resolution HST image
\citep{alc00b}. Towards the Galactic bulge, however, for which the
distance is about 7 times smaller than the LMC, blending may be less
severe.

The effect of blending is discussed in detail by \citet{dis95},
including the effects of blending when inferring properties of
underlying populations through the statistical study of lensing events
\citep{woz97,dom00}. The effect of blending can be summarized as
follows: (i) Blending makes it more difficult to observe a microlensing
event since the observed amplification is smaller than the actual
amplification. This decreases the efficiency of the detection of a
microlensing event. (ii) Blending enables some stars which are otherwise
invisible in the sample to be included in the sample of the monitored
events. This increases the efficiency of detection. (iii) The effect of
blending is to increase the number of monitored stars, thus increasing
the net efficiency of detection.

The first effect tends to offset the latter two. The net effect depends
on the brightness of the source/lens, and the crowding of the field. So
far as our analysis of extinction is concerned, since the blending star
is expected to be preferentially closer than the microlensed source,
blending dilutes the effect of extinction. As \citet{dis95} point
out, the effect of blending is more important for fainter sources. A
full analysis of blending is beyond the scope of this work. However,
our preliminary estimate suggests that, if the microlensed star is
brighter than $m_v \sim 18$, the contribution of the blended star is
about 10\%. This contribution increases to more than 50\% for stars
with $m_V \sim 20$. Thus, if the analysis is confined to brighter
sources (as is the case for the sources presented in \citet{kan03}),
the effect of blending can be neglected in the analysis of the
spectroscopic observations.

\section{EFFECTS OF EXTINCTION ON SPECTRA}

It has been shown that a predominant fraction of microlensed sources
are located on the far side of the Galactic bulge. This can be
observationally confirmed by observing the extinction effects in the
spectra of the lensed stars which, in turn, can be used to estimate
the fractional contributions of the disk and bulge stars to the
total microlensing optical depth.

To simulate the effects of extinction on various spectral types, the
spectral models used were those from the Kurucz database (Dr. R.
Kurucz, CD-ROM No. 13 \citep{kur93}). Each of the Kurucz models used
have been normalized for a solar metallicity and a distance of $D_s =
9.0 \, \mathrm{kpc}$. Each of the model spectra are recalculated for
$E_{B-V}$ values of 0.0, 0.2, 0.4, 0.6, 0.8 where the extinction
corrected spectrum, $s_c$, is calculated from the raw spectrum, $s_r$,
via
\begin{equation}
  (s_c)_\lambda = (s_r)_\lambda \times 10^{-(E_{B-V} \times e_g)/2.5} 
\end{equation}
where 
\begin{equation}
  e_g = \frac{A_\lambda}{E_{B-V}}
\end{equation}
The data for the Galactic extinction, $e_g$, was taken from \citet{sea79}.

Table 4 shows the stellar parameters used for the model spectra,
where $T_{eff}$ is the effective temperature, $\log g$ is the log
gravity, and $m_V$ is the apparent magnitude of the star. Then
varying $E_{B-V}$ from 0.4 to 0.8 is fairly representative of the
levels of extinction that exist within Baade's window, suggested by
\citet{sta96} to be in the range of 1.26 to 2.79 magnitudes.

Figures 8, 9, 10, and 11 show the effect of extinction on observed
spectra of various types of stars given in Table 4. There is a
significant change in spectral features, such as slope and line
strength, from an $E_{B-V}$ value of 0.0 to 0.8. The slope and the
features can be used to quantitatively estimate the extinction of the
microlensed stars in comparison to the general sample in the Galactic
bulge. It is apparent from the model spectra that, although blue stars
in the main sequence will always be brighter than red main-sequence
stars at all wavelengths, the effects of extinction will cause the
stellar population at the far side of the bulge to discriminate
against blue stars as source stars in microlensing.

From our previous arguments, we expect a majority of the microlensed
sources to show extinction which would correspond to color excess
($E_{B-V}$) values in the range 0.4 to 0.8. Clearly, a collection of
microlensed source spectra bearing this characteristic would become
statistically significant when estimating the contribution of bulge-bulge
lensing to the microlensing optical depth. If this is shown to be
statistically the case then this simple method can be used as a
statistical distance indicator for microlensed sources.

A knowledge of the distance to the source enables the distance to the
lens to be estimated. To demonstrate this, the Galactic bulge model
outlined in Section 1.3 shall be used. For a source located at a
distance $D_s$, the fractional area covered by the Einstein rings of
the intervening stars at a depth $dD_d$ at a distance $D_d$ is
\begin{equation}
  A_f(D_d) \, dD_d = \frac{4 \pi G}{c^2} \rho_0 \exp \left[ - \frac{|R_0 -
      D_d|}{i_0} \right] \frac{D_d (D_s-D_d)}{D_s} \, dD_d
\end{equation}

As discussed earlier, the probability of the source being microlensed is
highest when the source is at the far side of the bulge. The probability
distribution of the location of the lenses (in units of the distance to
the source) is simply given by $A_f (D_d)$, which is shown in Figure 12.
The figure shows that the probability peaks at a distance of $D_d = 0.85
D_s$ which, for $D_s = 9.0 \, \mathrm{kpc}$, corresponds to $D_d = 7.8 \,
\mathrm{kpc}$. It is worth noting that this does not imply that most of
the lenses are at $D_d = 7.8 \, \mathrm{kpc}$. This only implies that if
the source is at $D_s = 9 \, \mathrm{kpc}$, a lens distance of 0.85 $D_s$
is twice more likely than a lens distance of 0.6 $D_s$.

Since the angular Einstein ring radius is known from the characteristic
time scale of the event, the extinction exhibited in the spectra of
microlensed sources can be used as a means to estimate the size of the
Einstein ring radius in terms of AU. Detectable deviations in
microlensing light curves due to planetary masses depend highly upon the
star-planet distance in the lensing system \citep{ben96,gau00}, making this
method a useful technique to constrain calculations of planet detection
efficiencies.

\section{CONCLUSIONS}

Various Galactic models have been used to demonstrate the theoretical
effects of extinction upon the microlensing optical depth. The exponential
density model provides the best approximation of the microlensing optical
depth when taking extinction into account. For the exponential model, it
is shown that stars that are lensed within Baade's window are about 13
times more likely to be at the far side of the bulge than the near side.
The estimated optical depth is 
$\approx 4.50 \times 10^{-6}$ for $a_1 = 2.0$ and $a_2 = 3.0$,
which is comparable to the observed optical depths of $(3.3 \pm 1.2) \times
10^{-6}$ and $(3.9_{-1.2}^{+1.8}) \times 10^{-6}$ measured by \citet{uda94}
and \citet{alc97} respectively.

The calculations presented in this paper do not necessarily assume that
there is a onoe-to-one relation between distance and reddening. Rather we
assume that the sources will be more reddened than the lens stars and,
statistically, other non-microlensed sources in the same narrow field. The
follow-up study presented in \citet{kan03}, for example, selects
comparison stars which are very close to the microlensed sources so that
the angular distance between them, and hence the variation in foreground
extinction, is at a minimum.

The effects of extinction on spectra of microlensed sources have been
simulated using Kurucz model spectra. It is shown that the majority of
microlensed sources should exhibit extinction between $E_{B-V} = 0.4$
and $E_{B-V} = 0.8$. It is also shown that using extinction as a
distance indicator for microlensed sources may be used to statistically
estimate the distance to the lens from the probability distribution
corresponding to the appropriate Galactic model.

The measured extinctions for the microlensed sources can be used to
determine the expected stellar number density gradient $a$ which, in turn,
can be used to determine the optical depth more accurately. If there is a
smooth velocity gradient within the Galactic bulge, one would expect a
statistical correlation between the radial velocity and the extinction of
microlensed sources which, given enough samples, will provide useful
information regarding the 3-dimensional velocity structure of the far side
of the Galactic bulge. Given the variation in extinction through the
Galactic bulge, the sample would require $\sim 100$ spectra of microlensed
sources in order to clearly show the described effects.

\acknowledgements

The authors would like to thank Eric Agol for several useful discussions
regarding the Galactic bulge model calculations. Thanks are also due to
Rosanne Di Stefano for her useful comments regarding blending. We would
like to thank the referee Hongsheng Zhao for very useful comments which
led to an improvement of the paper.

\appendix

\section{FRACTIONAL AREA FOR EXPONENTIAL DENSITY}

This appendix details the evaluation of the integral in equation (26).
Let the integral be defined as
\begin{equation}
  I = \int_0^{D_s} \exp \left[ - \frac{|R_0 - D_d|}{i_0} \right]
  \frac{D_d (D_s-D_d)}{D_s} \, dD_d
\end{equation}
Due to the absolute value of $R_0 - D_d$, this integral is a boundary
value problem around the Galactocentric distance $R_0$. Hence, the
integral needs to be separated into two integrals defined on either of
the boundary $R_0$, as follows
\begin{displaymath}
  I_1 = \int_0^{D_s} \exp \left[ - \frac{(R_0 - D_d)}{i_0} \right]
  \frac{D_d (D_s-D_d)}{D_s} \, dD_d\\
\end{displaymath}
for $D_s < R_0$, and
\begin{eqnarray*}
  I_2 & = & \int_0^{R_0} \exp \left[ - \frac{(R_0 - D_d)}{i_0} \right]
  \frac{D_d (D_s-D_d)}{D_s} \, dD_d\\
  & & + \int_{R_0}^{D_s} \exp \left[ - \frac{(D_d - R_0)}{i_0} \right]
  \frac{D_d (D_s-D_d)}{D_s} \, dD_d
\end{eqnarray*}
for $D_s > R_0$. These two integrals will now be considered separately.

\subsection{Source Distance Less Than ${\bf R_0}$}

The integral $I_1$ may be expanded into two parts
\begin{equation}
  I_1 = \int_0^{D_s} D_d \exp \left[ - \frac{(R_0 - D_d)}{i_0} \right] \,
  dD_d - \int_0^{D_s} \frac{D_d^2}{D_s} \exp \left[ - \frac{(R_0 -
      D_d)}{i_0} \right] \, dD_d
\end{equation}
Both of these integrals may be evaluated using integration by parts:
\begin{displaymath}
  \int u \frac{dv}{dx} \, dx = uv - \int v \frac{du}{dx} \, dx
\end{displaymath}

In the case of the first integral, it is most convenient to define
\begin{displaymath}
  u = D_d \mbox{ and } \frac{dv}{dD_d} = \exp \left[ - \frac{(R_0 -
      D_d)}{i_0} \right]
\end{displaymath}
Then it follows that
\begin{displaymath}
  \frac{du}{D_d} = 1 \mbox{ and } v = \int \exp \left[ - \frac{(R_0 -
      D_d)}{i_0} \right] \, D_d
\end{displaymath}
By substituting $\gamma = (R_0 - D_d)/i_0$, $v$ becomes
\begin{eqnarray*}
  v & = & -i_0 \int e^{-\gamma} \, d\gamma\\
  & = & i_0 e^{-\gamma}\\
  & = & i_0 \exp \left[ - \frac{(R_0 - D_d)}{i_0} \right]
\end{eqnarray*}
Then using integration by parts
\begin{eqnarray}
  \int u \frac{dv}{dD_d} \, dD_d & = & D_d i_0 \exp \left[ - \frac{(R_0
      - D_d)}{i_0} \right] - i_0 \int \exp \left[ - \frac{(R_0 - D_d)}{i_0}
    \right] \, dD_d \nonumber\\
  & = & i_0 D_d \exp \left[ - \frac{(R_0 - D_d)}{i_0} \right] - i_0^2
  \exp \left[ - \frac{(R_0 - D_d)}{i_0} \right] \nonumber\\
  & = & i_0 \exp \left[ - \frac{(R_0 - D_d)}{i_0} \right] (D_d - i_0)
\end{eqnarray}

In the case of the second integral, it is convenient to define
\begin{displaymath}
  u = \frac{D_d^2}{D_s} \mbox{ and } \frac{dv}{dD_d} = \exp \left[
    - \frac{(R_0 - D_d)}{i_0} \right]
\end{displaymath}
Then it follows that
\begin{displaymath}
  \frac{du}{D_d} = \frac{2 D_d}{D_s} \mbox{ and } v = i_0 \exp
  \left[ - \frac{(R_0 - D_d)}{i_0} \right]
\end{displaymath}
Then using integration by parts
\begin{eqnarray}
  \int u \frac{dv}{dD_d} \, dD_d & = & \frac{D_d^2}{D_s} i_0 \exp
  \left[ - \frac{(R_0 - D_d)}{i_0} \right] - \frac{2 i_0}{D_s} \int
  D_d \exp \left[ - \frac{(R_0 - D_d)}{i_0} \right] \, dD_d \nonumber\\
  & = & i_0 \frac{D_d^2}{D_s} \exp \left[ - \frac{(R_0 - D_d)}{i_0}
    \right] - \frac{2 i_0^2}{D_s} \exp \left[ - \frac{(R_0 - D_d)}{i_0}
    \right] (D_d - i_0) \nonumber\\
  & = & \frac{i_0}{D_s} \exp \left[ - \frac{(R_0 - D_d)}{i_0} \right]
  (D_d^2 - 2 i_0 D_d + 2 i_0^2)
\end{eqnarray}

Combining Equation A.3 and Equation A.4 yields
\begin{eqnarray}
  I_1 & = & \left. i_0 \exp \left[ - \frac{(R_0 - D_d)}{i_0} \right] \left(
  D_d - i_0 - \frac{D_d^2}{D_s} + \frac{2 i_0 D_d}{D_s} -
  \frac{2 i_0^2}{D_s} \right) \right|_{D_d = 0}^{D_d = D_s} \\
  & = & i_0^2 \exp \left[ - \frac{(R_0 - D_s)}{i_0} \right] \left( 1 -
  \frac{2 i_0}{D_s} \right) + i_0^2 \exp \left[ - \frac{R_0}{i_0} \right]
  \left( 1 + \frac{2 i_0}{D_s} \right) \nonumber\\
  & = & \frac{i_0^2 e^{-R_0/i_0} (D_s (1 + e^{D_s/i_0}) + 2 i_0 (1 -
    e^{D_s/i_0}))}{D_s}
\end{eqnarray}

\subsection{Source Distance Greater Than ${\bf R_0}$}

Since the first component of integral $I_2$ is identical to integral $I_1$
except for the limits, this component may be solved by simply substituting
the limits $0 \rightarrow R_0$ into Equation A.5.
\begin{eqnarray}
  I_{2.1} & = & \left. i_0 \exp \left[ - \frac{(R_0 - D_d)}{i_0} \right]
  \left( D_d - i_0 - \frac{D_d^2}{D_s} + \frac{2 i_0 D_d}{D_s} -
  \frac{2 i_0^2}{D_s} \right) \right|_{D_d = 0}^{D_d = R_0} \nonumber\\
  & = & i_0 \left( R_0 - i_0 - \frac{R_0^2}{D_s} + \frac{2 i_0 R_0}{D_s}
  - \frac{2 i_0^2}{D_s} \right) - i_0 \exp \left[ - \frac{R_0}{i_0} \right]
  \left( - i_0 - \frac{2 i_0^2}{D_s} \right) \nonumber\\
  & = & \frac{i_0 (R_0 D_s + 2 i_0 R_0 + i_0 e^{-R_0/i_0} (D_s + 2 i_0)
    - i_0 D_s - 2 i_0^2 - R_0^2)}{D_s}
\end{eqnarray}

As was the case for integral $I_1$, the second component of integral
$I_2$ may be solved by expanding the integral into two parts
\begin{displaymath}
  I_{2.2} = \int_{R_0}^{D_s} D_d \exp \left[ - \frac{(D_d - R_0)}{i_0}
    \right] \, dD_d - \int_{R_0}^{D_s} \frac{D_d^2}{D_s} \exp \left[ -
    \frac{(D_d - R_0)}{i_0} \right] \, dD_d
\end{displaymath}
Using integration by parts, the first integral becomes
\begin{equation}
  I_{2.2.1} = - i_0 \exp \left[ - \frac{(D_d - R_0)}{i_0} \right] (D_d +
  i_0)
\end{equation}
and the second integral becomes
\begin{equation}
  I_{2.2.2} = - \frac{i_0}{D_s} \exp \left[ - \frac{(D_d - R_0)}{i_0} \right]
  (D_d^2 + 2 i_0 D_d + 2 i_0^2)
\end{equation}
so that combining Equation A.8 and Equation A.9 yields
\begin{eqnarray}
  I_{2.2} & = & \left. i_0 \exp \left[ - \frac{(D_d - R_0)}{i_0} \right]
  \left( - D_d - i_0 + \frac{D_d^2}{D_s} + \frac{2 i_0 D_d}{D_s} +
  \frac{2 i_0^2}{D_s} \right) \right|_{D_d = R_0}^{D_d = D_s} \nonumber\\
  & = & i_0 \exp \left[ - \frac{(D_s - R_0)}{i_0} \right] \left( i_0 +
  \frac{2 i_0^2}{D_s} \right) - i_0 \left( - R_0 - i_0 + \frac{R_0^2}{D_s}
  + \frac{2 i_0 R_0}{D_s} + \frac{2 i_0^2}{D_s} \right) \nonumber\\
  & = & \frac{i_0 (i_0 D_s - 2 i_0^2 + i_0 e^{-(D_s-R_0)/i_0} (D_s + 2 i_0)
    + R_0 D_s - 2 i_0 R_0 - R_0^2)}{D_s}
\end{eqnarray}

Combining Equation A.7 and Equation A.10 leads to the expression for
integral $I_2$
\begin{equation}
  I_2 = \frac{i_0 \left( 2 i_0^2 \left( e^{- \frac{D_s-R_0}{i_0}} + e^{-
      \frac{R_0}{i_0}} - 2 \right) - 2 R_0^2 + D_s \left( i_0 e^{-
      \frac{D_s-R_0}{i_0}} + i_0 e^{- \frac{R_0}{i_0}} + 2 R_0 \right)
    \right)}{D_s}
\end{equation}

\clearpage

\begin{deluxetable}{lrrr}
\tablecaption{Variation of instantaneous probability of microlensing
with extinction $a_c$ for a constant density between the observer and
the source.}
\tablewidth{0pt}
\tablehead{
\colhead{$a_c$} & \colhead{$A_f(7) N_{obs}(7) / N_{tot} (10^{-7})$} &
\colhead{$A_f(8) N_{obs}(8) / N_{tot} (10^{-7})$} &
\colhead{$A_f(9) N_{obs}(9) / N_{tot} (10^{-7})$}
}
\startdata
1.0 & 10.92 & 14.26 & 18.04\\
2.0 & 8.83 & 10.68 & 12.51\\
3.0 & 7.66 & 8.85 & 9.91\\
4.0 & 6.87 & 7.69 & 8.34\\
5.0 & 6.28 & 6.86 & 7.26\\
6.0 & 5.83 & 6.24 & 6.47\\
7.0 & 5.46 & 5.74 & 5.85\\
\enddata
\end{deluxetable}

\clearpage

\begin{deluxetable}{lrrr}
\tablecaption{Variation of instantaneous probability of microlensing
with extinction $a_c$ for a constant density for the disk and bulge.}
\tablewidth{0pt}
\tablehead{
\colhead{$a_c$} & \colhead{$A_f(7) N_{obs}(7) / N_{tot} (10^{-7})$} &
\colhead{$A_f(8) N_{obs}(8) / N_{tot} (10^{-7})$} &
\colhead{$A_f(9) N_{obs}(9) / N_{tot} (10^{-7})$}
}
\startdata
1.0 & 3.54 & 5.80 & 10.22\\ 
2.0 & 4.91 & 4.02 & 3.54\\
3.0 & 5.83 & 3.19 & 1.87\\
4.0 & 6.54 & 2.68 & 1.18\\
5.0 & 7.12 & 2.33 & 0.82\\
6.0 & 7.61 & 2.08 & 0.61\\
7.0 & 8.03 & 1.88 & 0.47\\
\enddata
\end{deluxetable}

\clearpage

\begin{deluxetable}{lrrr}
\tablecaption{Variation of instantaneous probability of microlensing
with extinction $a_2$ for an exponential density model and $a_1 = 1$.}
\tablewidth{0pt}
\tablehead{
\colhead{$a_2$} & \colhead{$A_f(7) N_{obs}(7) / N_{tot} (10^{-6})$} &
\colhead{$A_f(8) N_{obs}(8) / N_{tot} (10^{-6})$} &
\colhead{$A_f(9) N_{obs}(9) / N_{tot} (10^{-6})$}
}
\startdata
1.0 & 0.48 & 2.02 & 6.07\\
2.0 & 0.56 & 2.35 & 3.52\\
3.0 & 0.60 & 2.51 & 2.52\\
4.0 & 0.62 & 2.62 & 1.97\\
5.0 & 0.64 & 2.70 & 1.62\\
6.0 & 0.65 & 2.76 & 1.38\\
7.0 & 0.66 & 2.80 & 1.20\\
\enddata
\end{deluxetable}

\clearpage

\begin{deluxetable}{lrrr}
\tablecaption{Characteristics of example spectral types}
\tablewidth{0pt}
\tablehead{
\colhead{Spectral Type} & \colhead{$T_{eff}$} & \colhead{$\log g$} &
\colhead{$m_V$}
}
\startdata
M0III & 3800 & +1.34 & 14.4\\
G0V & 6030 & +4.39 & 19.2\\
F0V & 7200 & +4.34 & 17.4\\
B5III & 15000 & +3.49 & 12.4\\
\enddata
\end{deluxetable}

\clearpage

\begin{figure}
\plotone{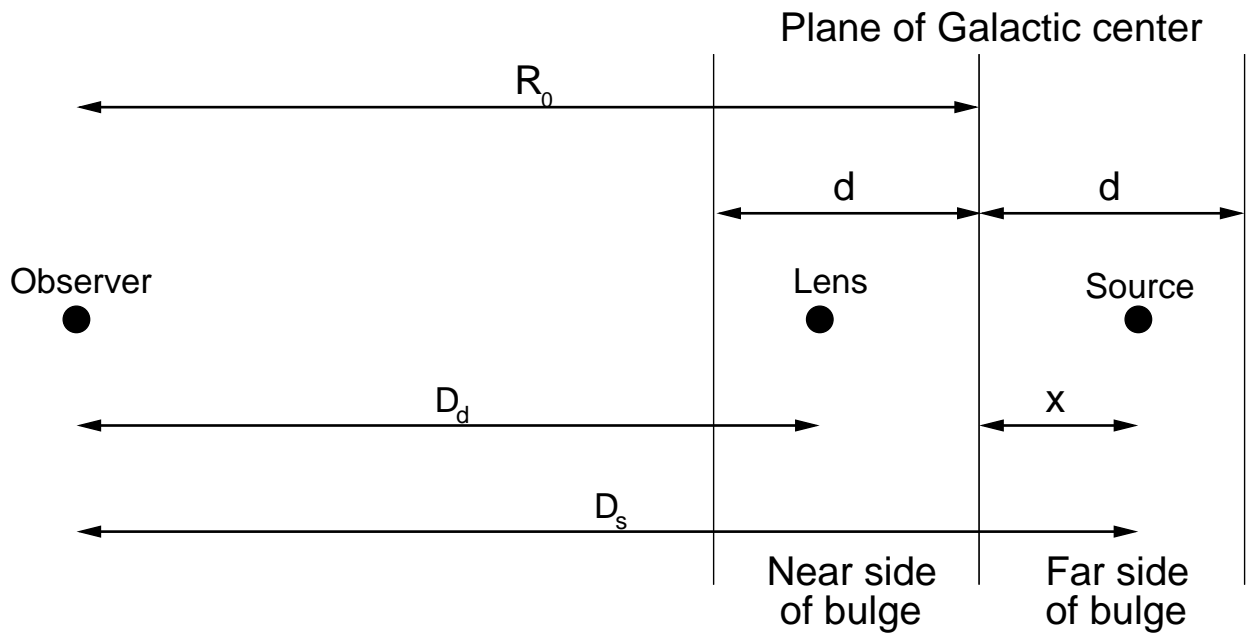}
\caption{Schematic of the geometry used in the Galactic model assuming
that all the microlensed sources are within the bulge.}
\end{figure}

\begin{figure}
\plotone{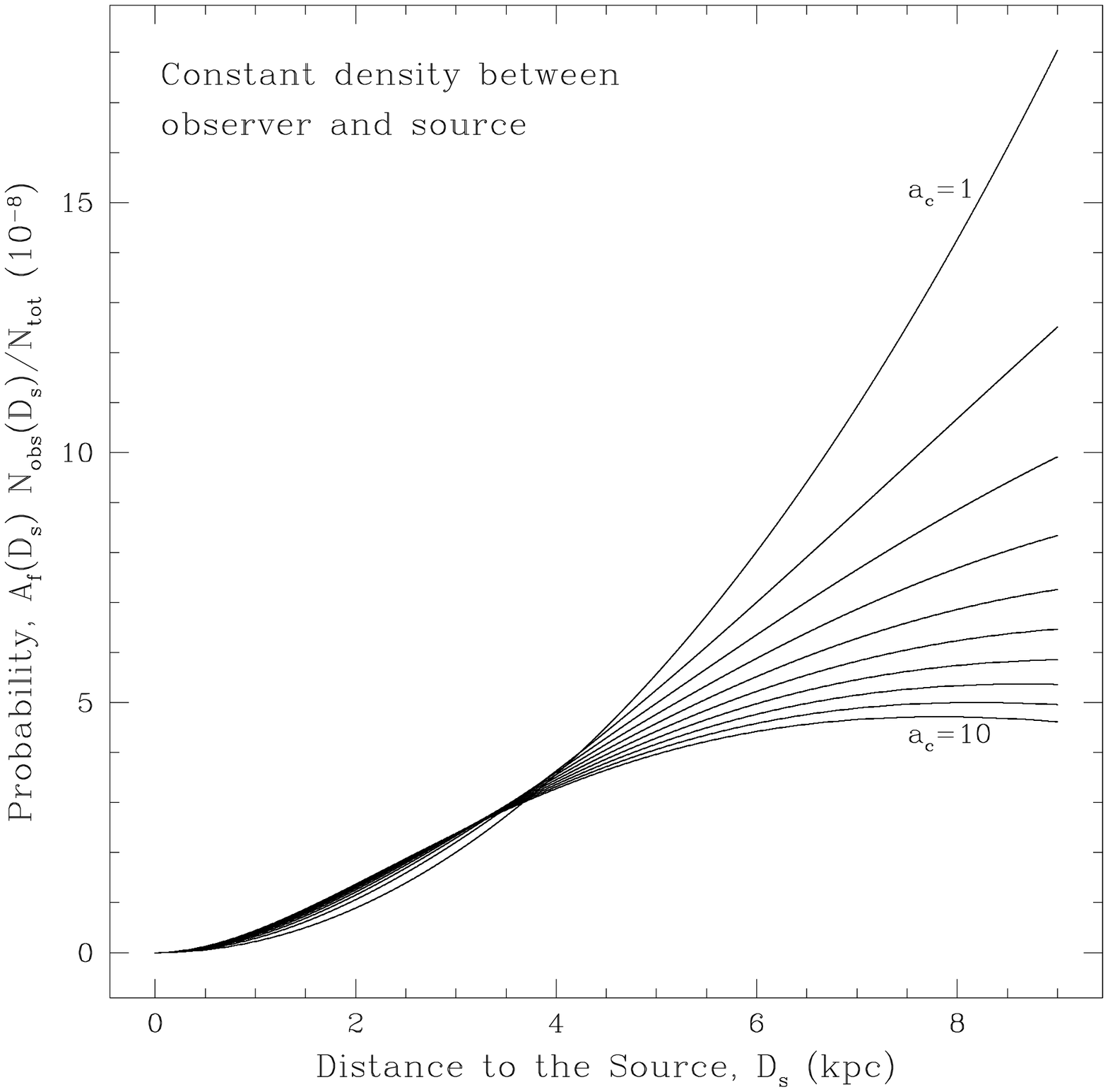}
\caption{Probability that a given microlensed source is at a distance
$D_s$ as a function of $D_s$, for values of $a_c$ ranging from $a_c = 1$
(uniform density of observed stars) to $a_c = 10$.}
\end{figure}

\begin{figure}
\plotone{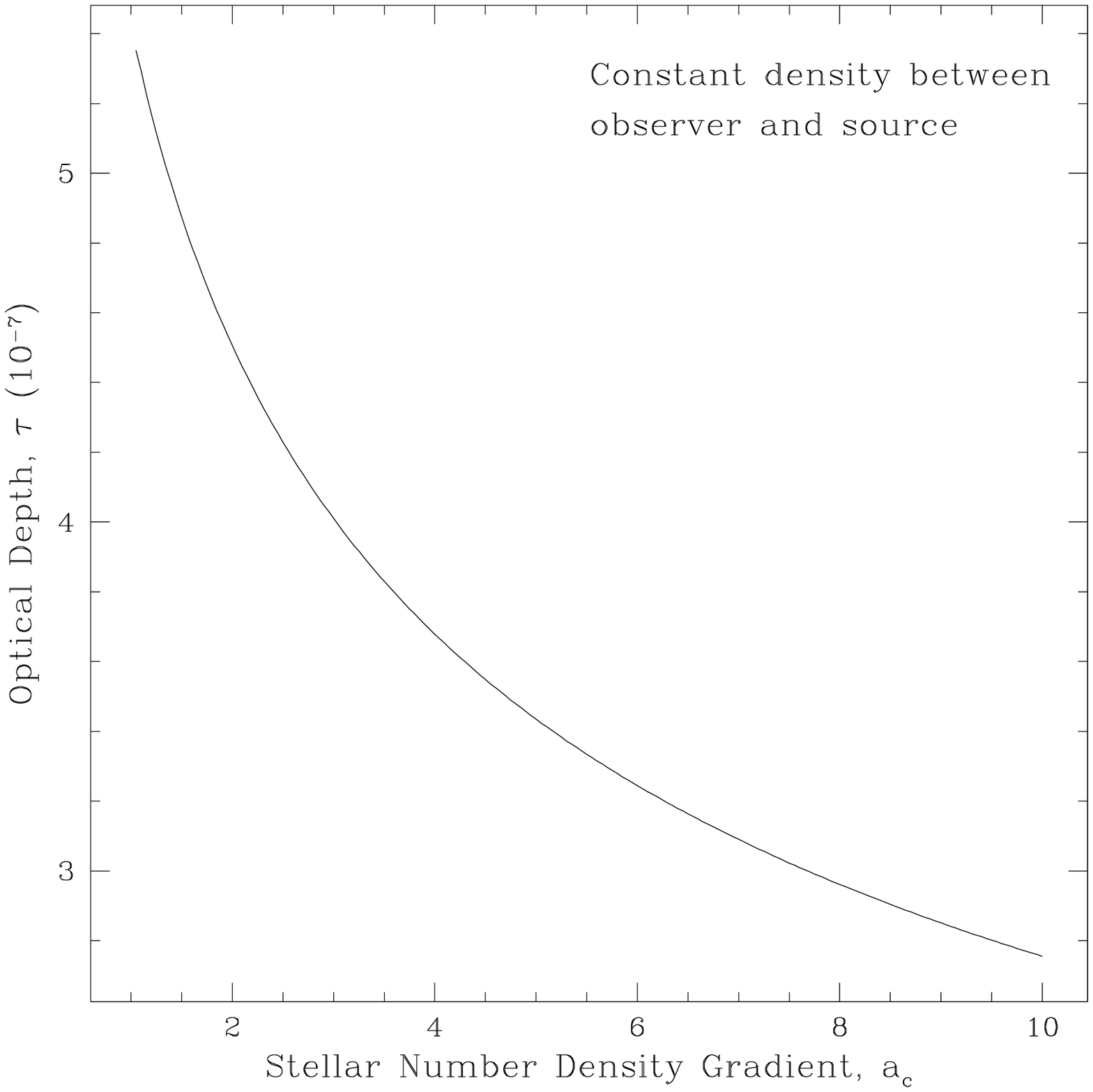}
\caption{Optical depth as a function of $a_c$, which is the ratio of the
maximum to the minimum observed stellar number density in a given line
of sight, assuming a constant matter density for the lenses.}
\end{figure}

\begin{figure}
\plotone{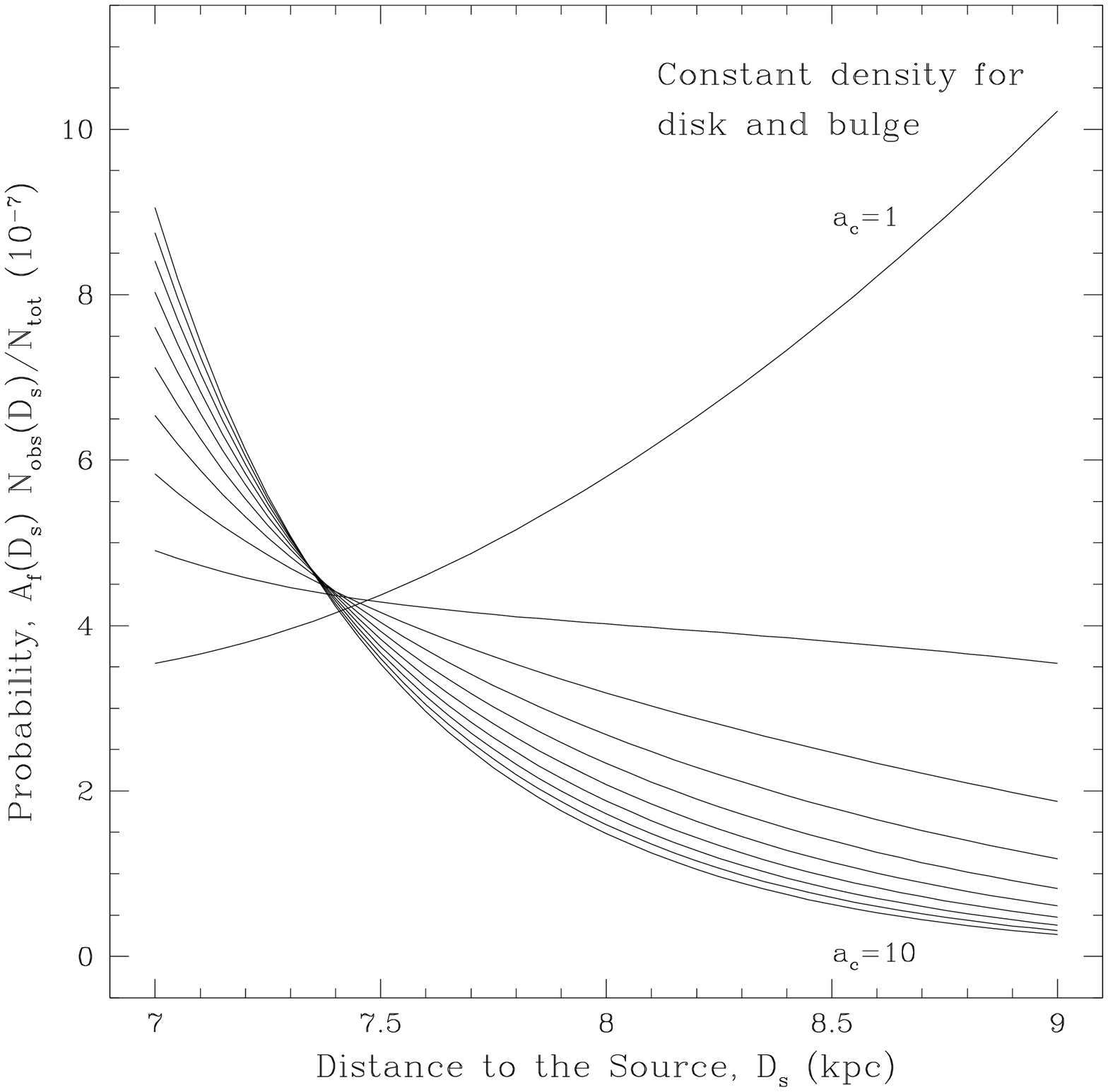}
\caption{Probability that a given microlensed source is at a distance
$D_s$ as a function of $D_s$, for values of $a_c$ ranging from $a_c = 1$
(uniform density of observed stars) to $a_c = 10$.}
\end{figure}

\begin{figure}
\plotone{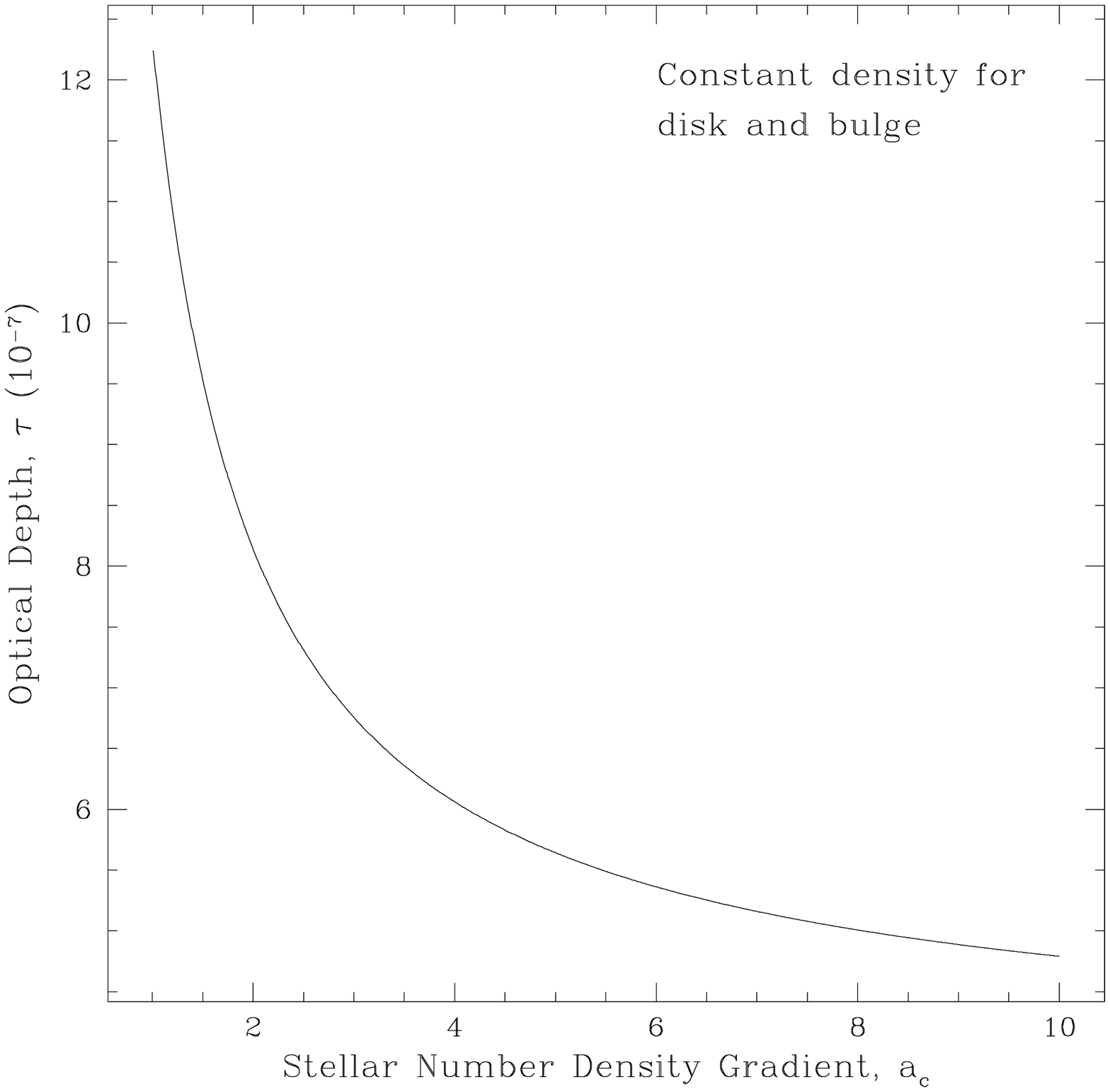}
\caption{Optical depth as a function of $a_c$, which is the ratio of the
maximum to the minimum observed stellar number density in a given line
of sight, assuming that the matter density for the lenses is constant in
the disk and in the bulge.}
\end{figure}

\begin{figure}
\plotone{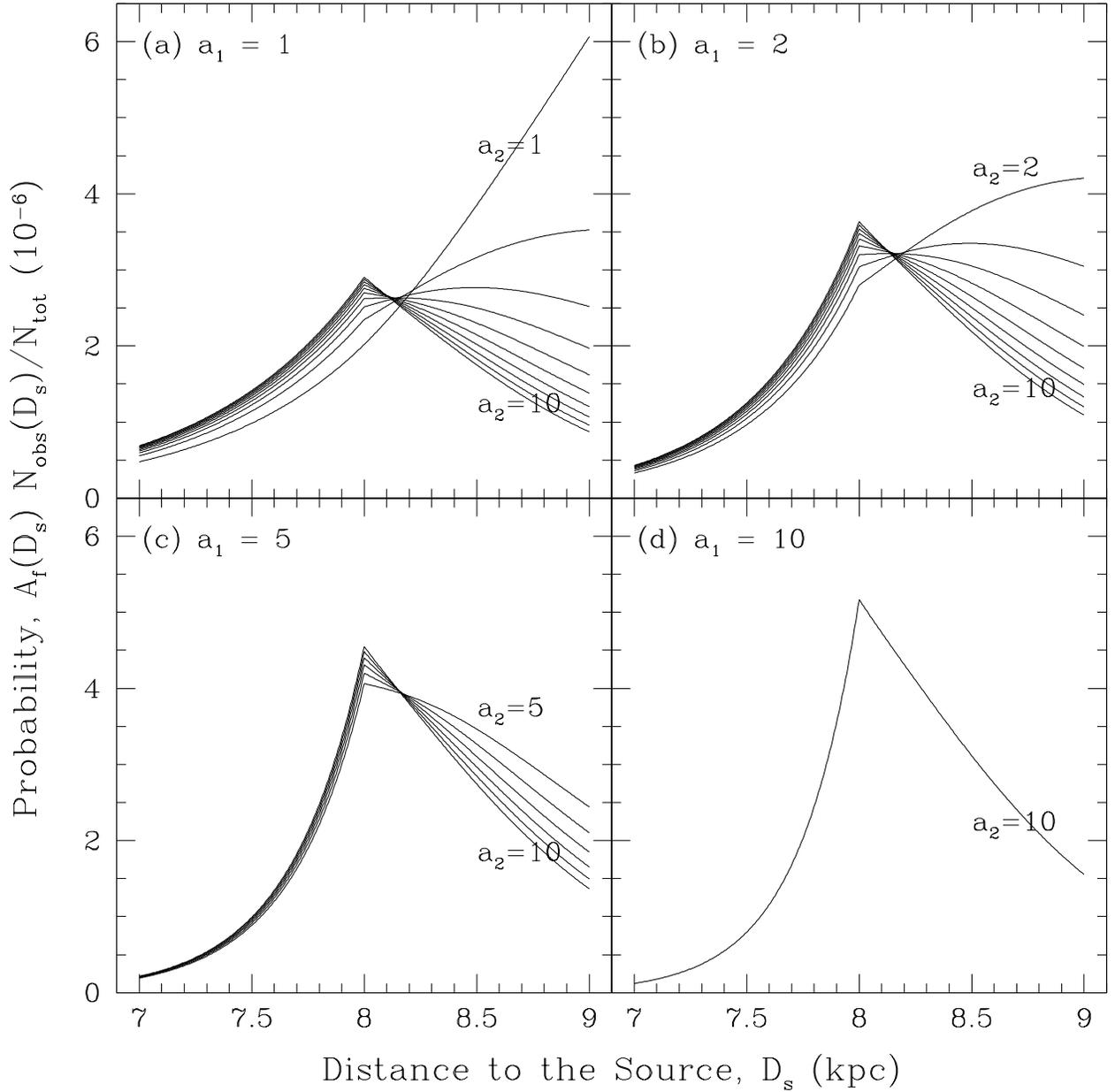}
\caption{Probability that a given microlensed source is at a distance
$D_s$ as a function of $D_s$, for values of $a_2$ ranging from $a_2 = a_1$
to $a_2 = 10$. This has been calculated for (a) $a_1 = 1$, (b) $a_1 = 2$,
(c) $a_1 = 5$, and (d) $a_1 = 10$ with the restriction as expected from
physical grounds that $a_1 < a_2$.}
\end{figure}

\begin{figure}
\plotone{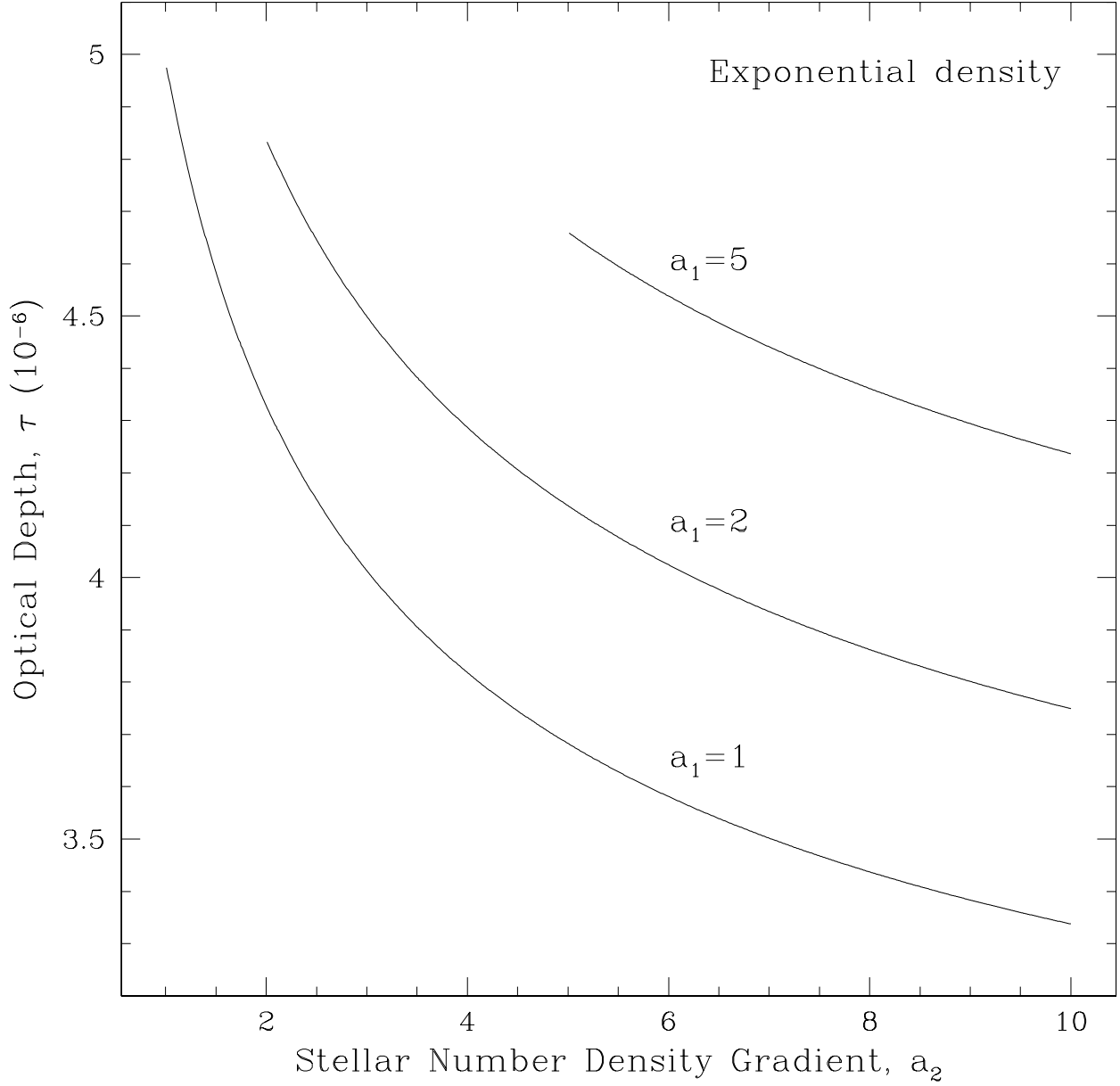}
\caption{Optical depth as a function of $a_2$, which is the ratio of the
maximum to the minimum observed stellar number density in a given line
of sight, assuming an exponential matter density for the lenses. This has
been calculated for $a_1 = 1$, $a_1 = 2$, and $a_1 = 5$ with the
restriction as expected from physical grounds that $a_1 < a_2$.}
\end{figure}

\begin{figure}
\rotatebox{-90}{\plotone{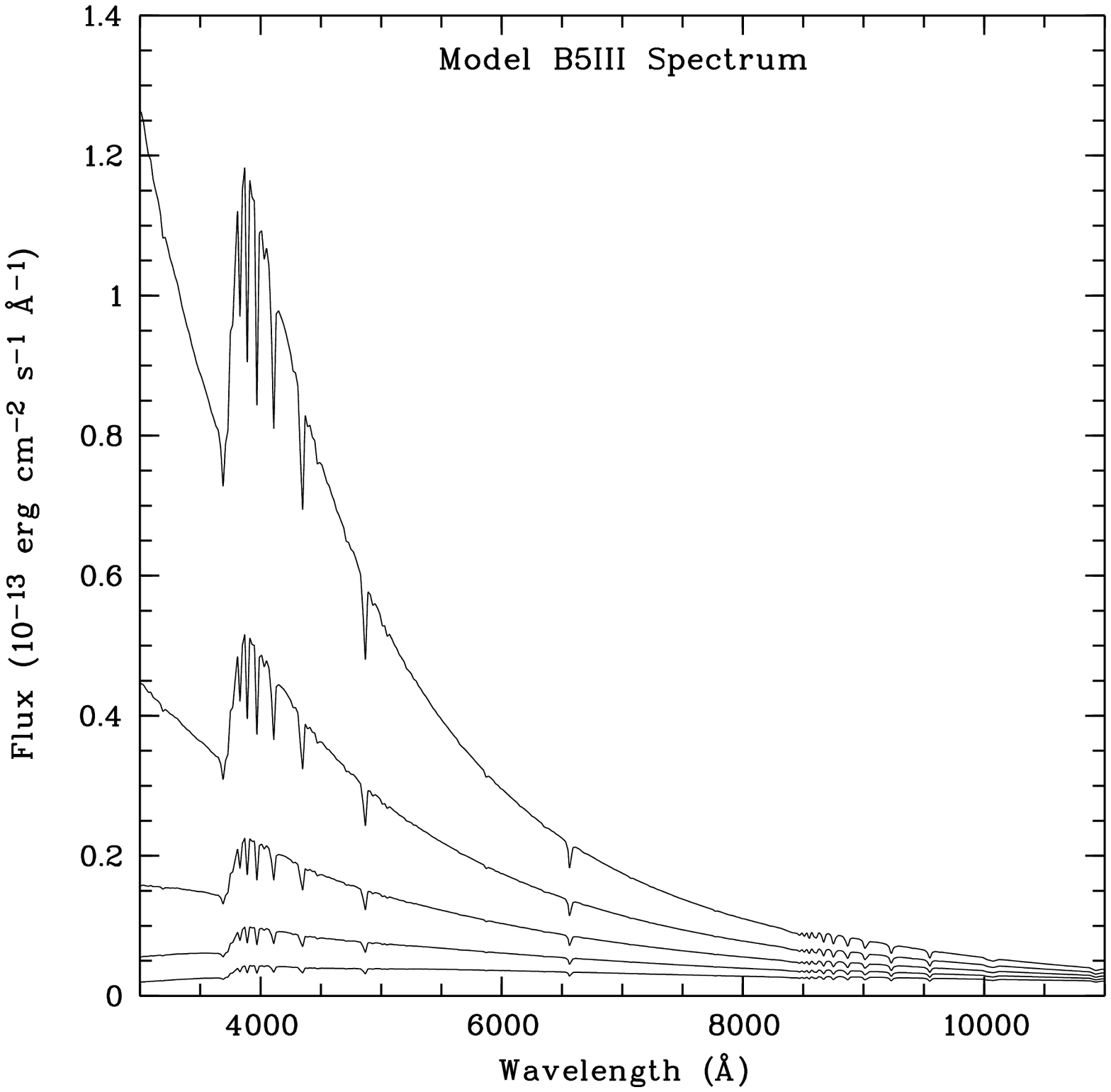}}
\caption{Model spectra of a spectral type B5III star showing
the effect of increasing levels of extinction corresponding to 
$E_{B-V}$ of 0.0 to 0.8 in steps of 0.2.}
\end{figure}

\begin{figure}
\rotatebox{-90}{\plotone{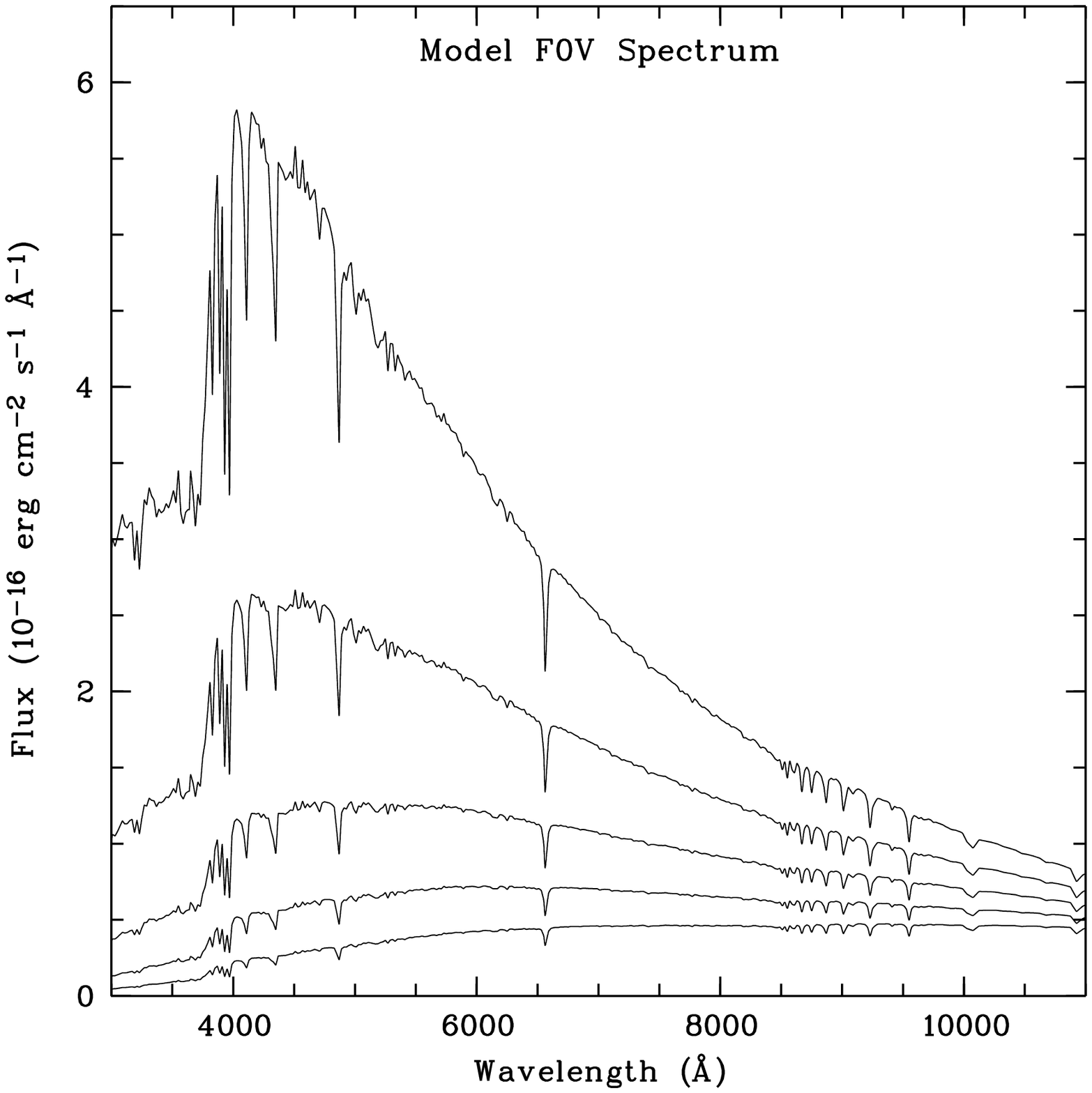}}
\caption{Model spectra of a spectral type F0V star showing the
effect of increasing levels of extinction corresponding to
$E_{B-V}$ of 0.0 to 0.8 in steps of 0.2.}
\end{figure}

\begin{figure}
\rotatebox{-90}{\plotone{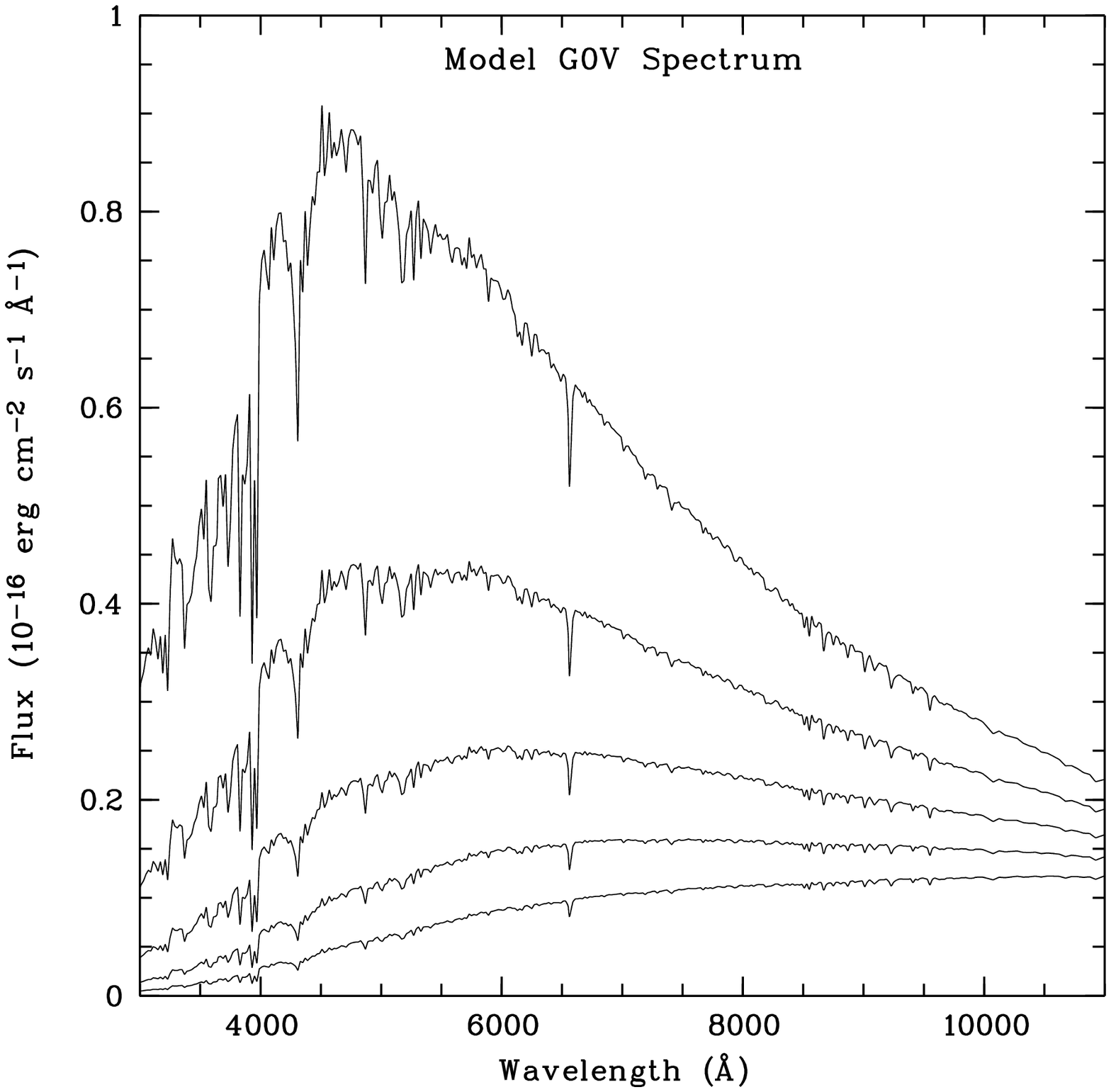}}
\caption{Model spectra of a spectral type G0V star showing the
effect of increasing levels of extinction corresponding to
$E_{B-V}$ of 0.0 to 0.8 in steps of 0.2.}
\end{figure}

\begin{figure}
\rotatebox{-90}{\plotone{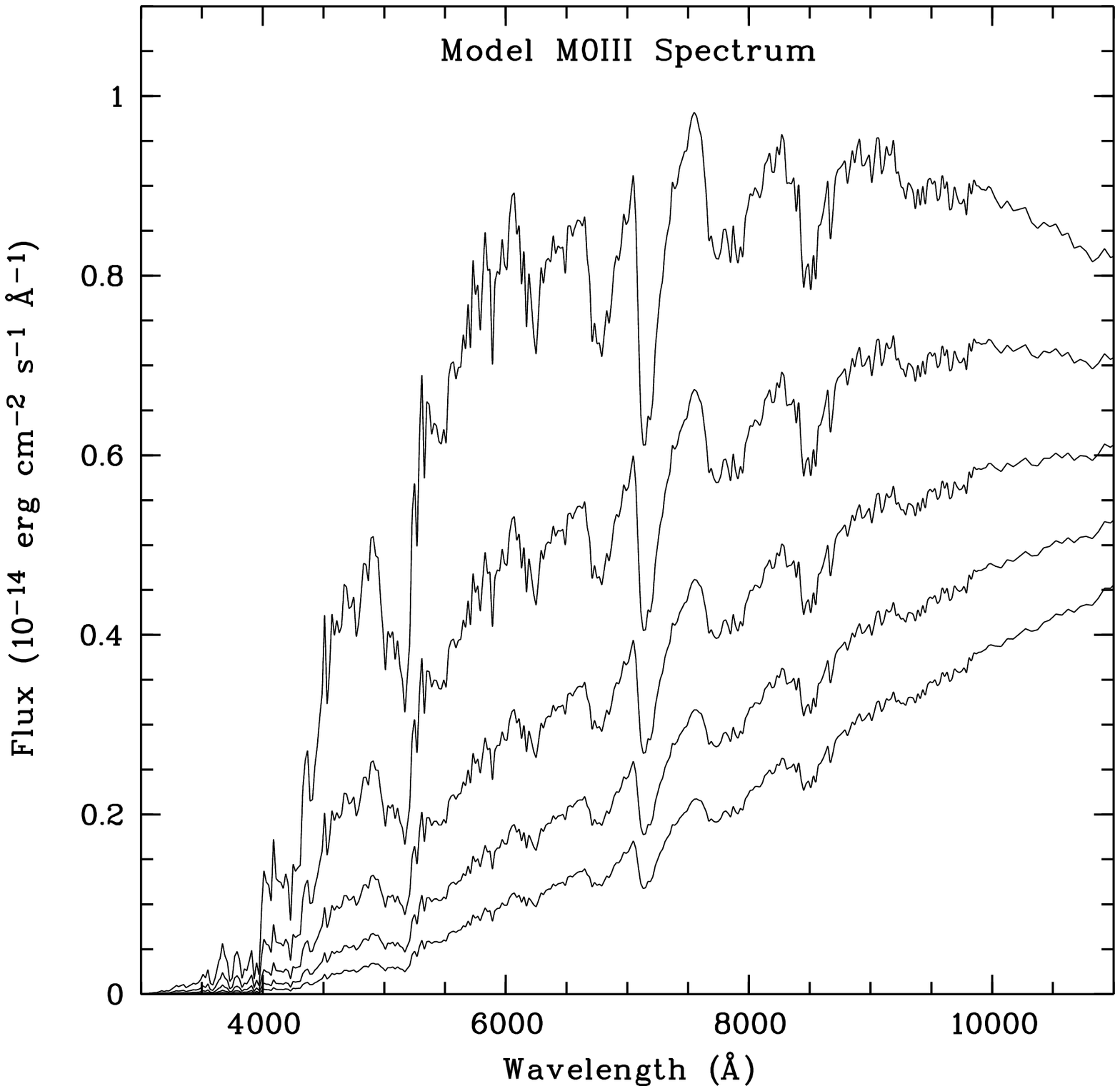}}
\caption{Model spectra of a spectral type M0III star showing
the effect of increasing levels of extinction corresponding to 
$E_{B-V}$ of 0.0 to 0.8 in steps of 0.2.}
\end{figure}

\begin{figure}
\plotone{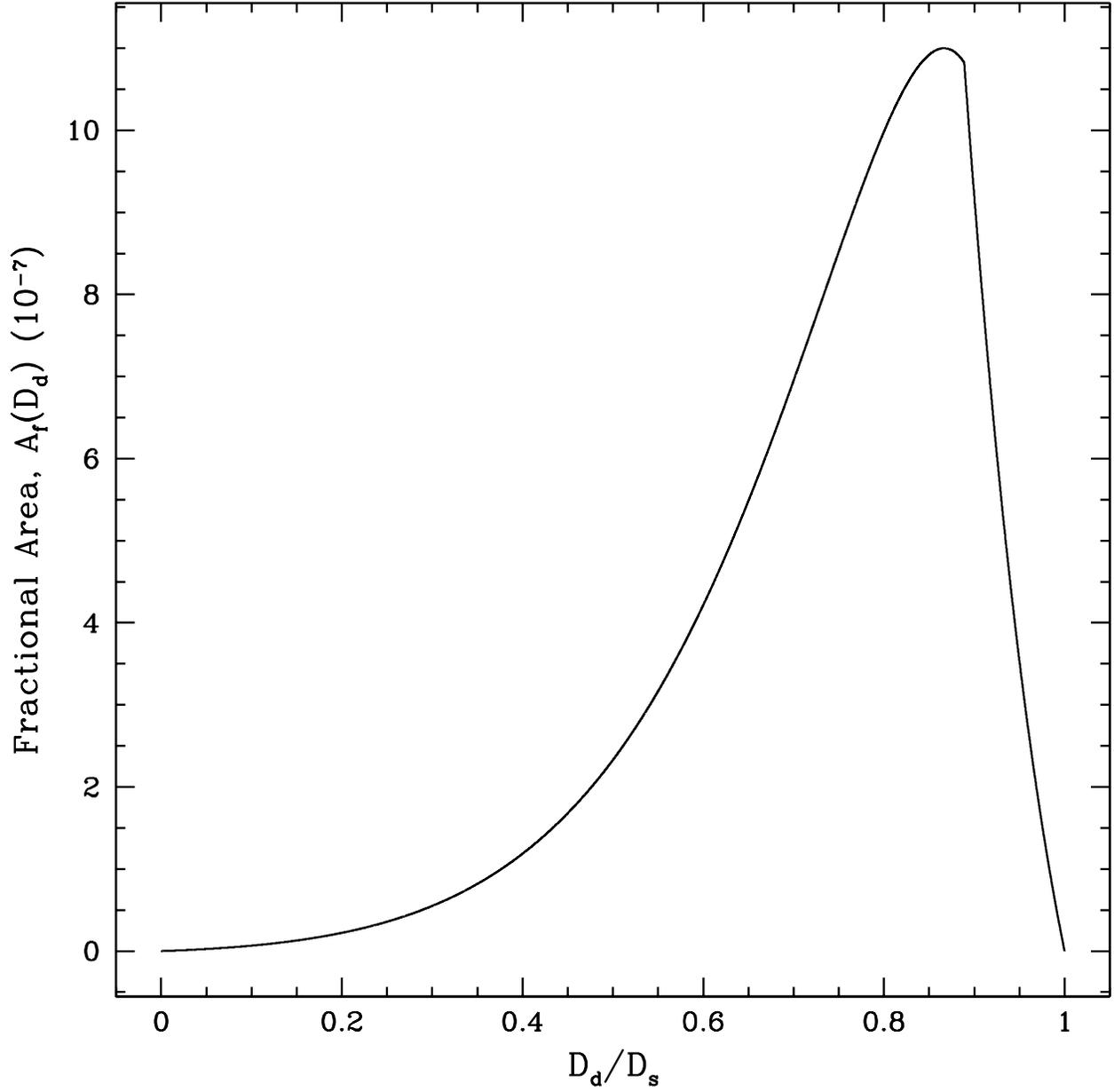}
\caption{A probability distribution for lenses given a source
located at a distance of $D_s = 9.0 \, \mathrm{kpc}$. In this case, it is
most probable that the lens will be located at a distance of $D_d = 7.8 \,
\mathrm{kpc}$.}
\end{figure}

\end{document}